\documentclass[runningheads]{llncs}
\usepackage[T1]{fontenc}
\usepackage{appendix}
\usepackage{graphicx}

\usepackage{bbm}
\usepackage{amsmath}
\usepackage{amssymb}
\usepackage{amsfonts}
\usepackage{color}
\usepackage{booktabs}

\usepackage[switch]{lineno}
\usepackage{adjustbox}
\usepackage[misc]{ifsym}
\newcommand{\yp}{\textcolor{black}}

\usepackage{graphicx}%
\usepackage{multirow}%
\usepackage{mathrsfs}%
\usepackage{manyfoot}%
\usepackage{booktabs}%
\usepackage{algorithm}%
\usepackage{algorithmicx}%
\usepackage{algpseudocode}%
\usepackage{listings}%
\usepackage{hyperref}

\begin{document}
\title{PMT-IQA: Progressive Multi-Task Learning for Blind Image Quality Assessment}

\titlerunning{PMT-IQA: Progressive Multi-Task Learning}
%
\author{Qingyi Pan\inst{1,2} \and
 Ning Guo\inst{1} \and
 Letu Qingge\inst{3} \and 
 Jingyi Zhang\inst{2} \and 
 Pei Yang\inst{1}$^{(\textrm{\Letter})}$
 }

%
\authorrunning{Q. Pan~et~al.}
%
 \institute{Department of Computer Technology and Application, Qinghai University, China \and
  Center for Statistical Science \& Department of
 Industrial Engineering, Tsinghua University, China \and
 Department of Computer Science, North Carolina A\&T State University, USA 
 }

\maketitle              
\begin{abstract}
Blind image quality assessment (BIQA) remains challenging due to the diverse types of distortion and variable image content, which complicates the distortion patterns crossing different scales and aggravates the difficulty of the regression problem for BIQA. However, existing BIQA methods often fail to consider multi-scale distortion patterns and image content, and there has limited research on improving the performance of quality regression models through specific learning strategies. In this paper, we propose a simple yet effective Progressive Multi-Task Image Quality Assessment (PMT-IQA) model, which contains a multi-scale feature extraction module (MS) and a progressive multi-task learning module (PMT), to help the model learn complex distortion patterns and better optimize the regression issue to align with the law of human learning process from easy to hard. To verify the effectiveness of the proposed PMT-IQA model, we conduct experiments on four widely used public datasets, and the experimental results indicate that the performance of PMT-IQA is superior to the comparison approaches, and both MS and PMT modules improve the model's performance. The source code for this study is available at \href{https://github.com/pqy000/PMT-IQA}{https://github.com/pqy000/PMT-IQA}.

\end{abstract}
\section{Introduction}
\label{sec:intro}

With the popularity of smartphones and other camera devices in recent years, a vast amount of images have been produced and play an increasingly important role in people's information interaction. However, these images could be distorted (i.e. quality degradation caused by noise, lossy compression, etc) by various factors, including the professional level of the photographer, equipment performance, transmission and device storage, etc. Therefore, it is of great need to assess the quality of images. Although people can subjectively evaluate the image quality accurately and reliably, it is very limited in practical applications due to time-consuming and laborious~\cite{zhai2020perceptual}. Consequently, \yp{objective image quality assessment} (IQA) \cite{kim2017deep},  which aims to explore models for automatically evaluating the image quality in line with the human vision system (HVS), has attracted much attention in the past few years~\cite{zhai2020perceptual} \cite{wang2004image}. Among all the objective IQA methods, blind IQA (BIQA), also called no-reference IQA (NR-IQA), approaches are the most challenging. ``Blind'' in this context refers to the fact that no pristine images (i.e. undistorted reference images) are required during the process of image quality evaluation. Yet much progress has been made on this topic, it is still an open and challenging issue, and in this study, we are committed to exploring the BIQA problem.

\begin{figure}[t]
    \centering
    \includegraphics[width=0.80\textwidth]{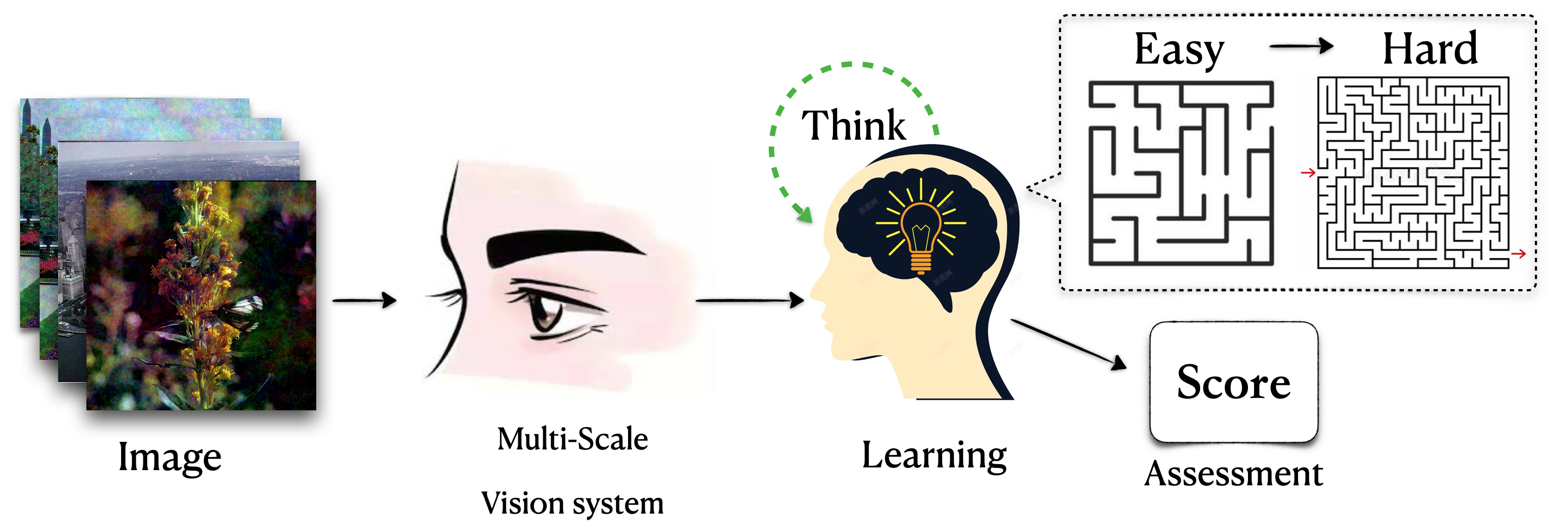}
    \caption{Motivation diagram of the proposed PMT-IQA. We divide image quality assessment into two parts, which are multi-scale vision system and human learning law based image quality assessment.}
    \label{fig::motivation}
\end{figure}

The diversity of distortion and variability in image content are the main reasons why BIQA is full of challenges. On the one hand, they complicate the distortion patterns, covering multiple scales, from local to global. On the other hand, the complex input space aggravates the difficulty of the regression problem for BIQA. However, existing works often fail to consider multi-scale distortion patterns and image content. Some attempts have been made to design end-to-end architectures for IQA. For example, Li~\textit{et al.} \cite{li2018has} extract features using a pre-trained deep convolutional neural network~(DCNN). It is evident that DCNN learns global features due to the gradual expansion of the receptive field in the convolutional layers as the network becomes deeper. However, most real-world image data distortion patterns exist in local areas. Therefore, the global features are not enough to capture the complex distortions. In addition, the human learning process follows the law from easy to hard, which is known as the easy-to-hard effect proposed by Pavlov~\cite{pavlov1927conditioned} in 1927. However, existing BIQA methods tend to solve the complex regression problem directly.

In this paper, we proposed a simple yet effective image quality assessment architecture inspired by the multi-scale characteristics of HVS and the from easy to hard law of human learning shown in Fig.~\ref{fig::motivation}. We name the proposed network as Progressive Multi-Task Image Quality Assessment~(PMT-IQA), since it is designed to capture distortion-related patterns using a task transfer strategy simulating the from easy to hard human learning law. The idea behind the proposed model is as below.
Firstly,  we extract global-to-local distortion-aware features by designing a multi-scale semantic feature extraction module. Secondly, inspired by the from easy to hard learning law, we build a progressive multi-task learning scheme, which can gradually shift from an easy task (i.e. quality level classification) to a hard one (quality score regression). At last, we evaluate the performance of the proposed PMT-IQA on several widely used public IQA datasets, and the experimental results validate the effectiveness of the PMT-IQA model.

\begin{figure*}[ht]
    \centering
    \includegraphics[width=0.98\textwidth]{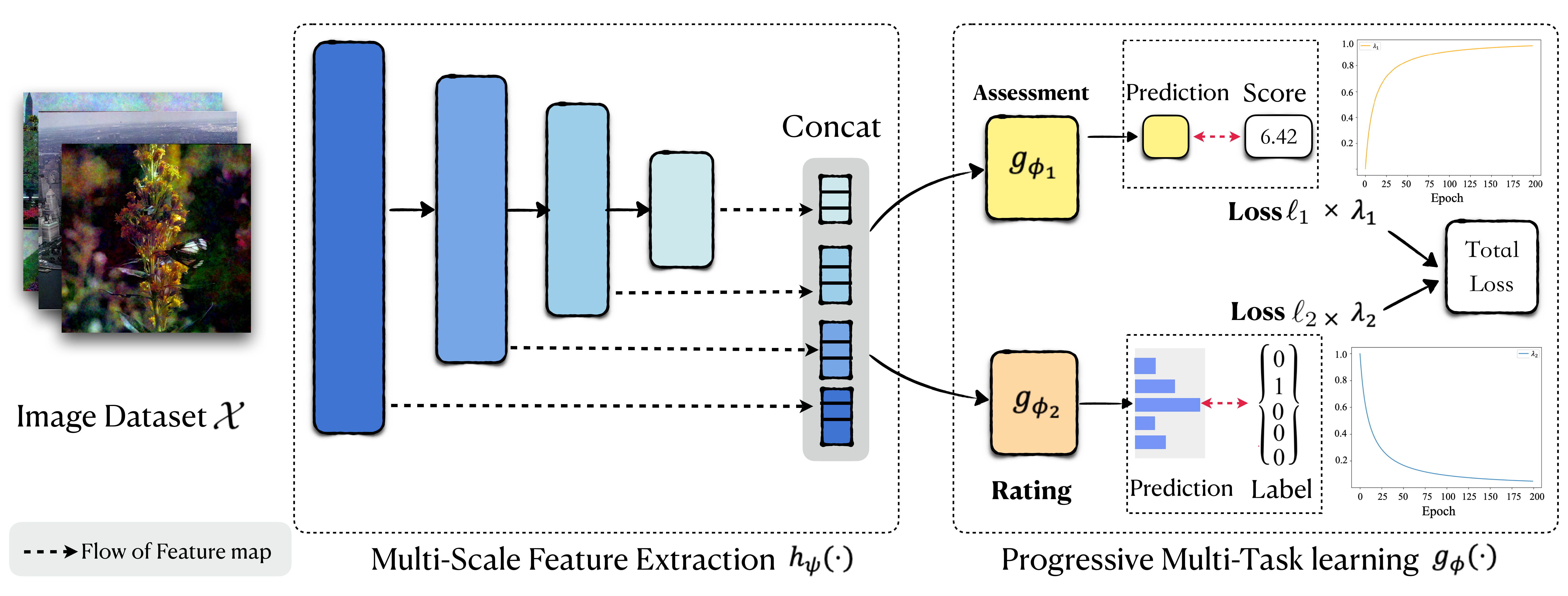}
    \caption{Progressive Multi-Task learning Image Quality Assessment architecture. It divides the task of IQA into two steps: Multi-Scale Semantic Feature Extraction and Progressive Multi-Task learning.}
    \label{fig::architecture}
\end{figure*}

\section{Related Works}
The development of deep neural network technology has greatly prompted the research of BIQA problem in the past few years. Zhang et~al.~\cite{zhang2018blind} proposed a dual-branch network that adapts to both authentic and synthetic distortions. 
Su et~al.~\cite{su2020blindly} introduced a self-adaptive hyper network for real-world distorted images to address the challenges of diverse distortion types and various contents in real-world images. To overcome the issue of weak cross-scenario capability of BIQA models, Zhang et~al.~\cite{zhang2021uncertainty} presented a unified uncertainty-aware BIQA model that cover both laboratory and wild scenarios. In addition, some latest deep learning technologies, including  contrastive learning~\cite{madhusudana2022image}, graph convolutional neural network~\cite{sun2022graphiqa}, and transformers~\cite{golestaneh2022no}, have also been used for BIQA research. While these models have achieved impressive performance, they mainly focus on network architecture design, but rarely study how to make the models learn better. Furthermore, these existing models also rarely used multi-scale for feature extraction.

Inspired by the easy-to-hard effect~\cite{pavlov1927conditioned}, we proposed a new learning scheme that imitates the law of from easy to hard learning for BIQA. In fact, curriculum learning~\cite{bengio2009curriculum}, which has received a lot of attention in the past decade~\cite{wang2021survey}, is a realization of this idea. It should be noted that curriculum learning simulates the from easy to hard learning law in terms of training data, which is to train a machine learning model from easier samples to hard ones. However, as the easy-to-hard effect indicates  \textit{``Early experience with an easy version of a discrimination task facilitates subsequent learning of a more difficult task ~\cite{pavlov1927conditioned}''}, our idea is to simulate the from easy to hard learning law through progressive learning of related tasks with different difficulty levels. We utilize the multi-task architecture in our implementation. Different from existing multi-task based BIQA models~\cite{kang2015simultaneous,ma2018end}, which simultaneously predict distortion type and image quality score, the two tasks (i.e. quality level prediction and quality score prediction) in our model are progressive. On the other hand, we employ dynamic task weights rather than fixed weights to transit from quality level classification to quality score regression. In summary, we proposed a novel progressive multi-task learning scheme simulating the from easy to hard learning law for BIQA.

\section{Methods}
\subsection{Overview of the Proposed Model}

The architecture of the proposed Progressive Mult-Task Image Quality Assessment~(PMT-IQA) model is presented in Fig.~\ref{fig::architecture}. It contains a multi-scale feature extraction module~(MS) and a progressive multi-task learning module~(PMT). The MS module is designed to extract image features with stronger representation ability by utilizing multi-scale information, and the PMT module is used to simulate the from easy to hard learning law through a dynamically weighted two-head multi-task structure. The proposed PMT-IQA model can be mathematically represented as Eq.~(\ref{eq::PMT-IQA}).
\begin{equation}
f_{\theta}(\cdot)=g_{\phi} \circ h_{\psi}(\cdot)
\label{eq::PMT-IQA}
\end{equation}
where $f_{\theta}(\cdot)$ represents the complete model with paramters $\theta$, $h_{\psi}$ is the MS module, which obtains local-to-global distortions, and PMT module $g_{\phi}$ learns complex regression problems. The operator $\circ$ in Eq.~(\ref{eq::PMT-IQA}) represents the composition of modules $g_{\phi}$ and $h_{\psi}$, where the output of $h_{\psi}$ serves as the input of $g_{\phi}$. The definition of the parameters $\theta=\{\phi, \psi\}$ will be declared in the next section.

\subsection{Multi-Scale Semantic Feature Extraction}
To characterize various distortions, we utilize convolutions to extract multi-scale features~(from local to global), each of which corresponds to a feature map $s_i$. Assuming we have features on \textit{n} scales, then we concatenate all features, as shown in Eq.~(\ref{eq::hidden}).
\begin{equation}
h_{\psi}(x_i) = {\rm concat}(s_1,\cdots s_j, \cdots, s_n)
\label{eq::hidden}
\end{equation}
More specifically, we use a pre-trained ResNet50~\cite{he2016deep} as the backbone architecture in PMT-IQA and collect feature maps from four stages of ResNet50.
Then we use $1\times 1$ convolution and global average pooling for dimension alignment.
The output of MS module $h(\cdot)$ is fed into the PMT module for prediction.

\subsection{Progressive Multi-Task Image Quality Assessment}
As introduced in section~\ref{sec:intro}, the diversity of distortion and variability of image content make the input space of quality scalar score regression issue complicated and increase the difficulty of model learning. To address this challenge, we propose a progressive multi-task learning scheme to mimic the from easy to hard human learning process. Specifically, in addition to the quality score prediction task, we introduce a relative simple quality-level classification task related to quality regression. During the model training process, we dynamically adjust the weights of the quality score prediction task and the quality-level classification task in the learning objective function, aiming to progressively shift the model's attention from the quality-level classification task to the quality score prediction task.

For the quality-level classification task, we divide the range of scalar quality scores into discrete sub-intervals, and let each sub-interval be a quality category, which represents a specific quality level, for the quality classification task. Let $w$ be the interval length, $[y^{min}, y^{max}]$ be the range of quality score, then we can obtain $K$ categories as:
\begin{equation}
K=\lfloor \frac{|y^{\max} - y^{\min}|}{w} \rfloor
\label{eq::discrete}
\end{equation}
For sample $x_i$ with scalar quality score $y_i$, we can get the corresponding quality category label $y_i^c \in  Y=\{1,\cdots,K\}$ by mapping $y_i$ into the corresponding discrete quality interval.

As shown in Fig.~\ref{fig::architecture}, the PMT $g_{\phi}$ contains two parts: \textit{scalar image quality score assessment} module $g_{\phi_1}:\mathbb{R}^{h}\rightarrow \mathbb{R}$ and \textit{image quality level classification} module $g_{\phi_2}:\mathbb{R}^{h}\rightarrow [0,1]^{K}$. Both $g_{\phi_1}$ and $g_{\phi_2}$ are implemented using a simple Multilayer Perceptron (MLP), where $g_{\phi_1}$ is composed of four fully connected layers and $g_{\phi_2}$ contains three fully connected layers and one softmax layer. ReLU() is utilized as the activation function of the first 3 and 2 fully connected layers of $g_{\phi_1}$ and $g_{\phi_2}$, respectively. Suppose $\phi_1=\{W^{\left(\phi_1\right)}_1, W^{\left(\phi_1\right)}_2, W^{\left(\phi_1\right)}_3, W^{\left(\phi_1\right)}_4\}$ and $\phi_2=\{W^{\left(\phi_2\right)}_1, W^{\left(\phi_2\right)}_2, W^{\left(\phi_2\right)}_3\}$, where $W^{\left(\phi_1\right)}_i$ and $W^{\left(\phi_2\right)}_i$
are the parameters of the $i$-th layer of $g_{\phi_1}$ and $g_{\phi_2}$ respectively, then for an input $X$ (note that $X$ is actually $[\hat{X};1]$ corresponding to real input $\hat{X}$ as $W_i^{\phi_j}$ represents weight and bias), $g_{\phi_1}$ and $g_{\phi_2}$ are defined as follows:
\begin{equation}
    g_{\phi_1}(X)=W^{(\phi_1)}_4(\mathrm{ReLU}(W^{(\phi_1)}_3(\mathrm{ReLU}(W^{(\phi_1)}_2(\mathrm{ReLU}(W^{(\phi_1)}_1 X ))))))
\end{equation}

\begin{equation}
    g_{\phi_2}(X) = (\frac{\exp(o_1)}{\sum_{i=1}^K\exp(o_i)}, \cdots, \frac{\exp(o_K)}{\sum_{i=1}^K\exp(o_i)})
\end{equation}
where $o_i$ is the \textit{i}-th component of the output of the last fully connected layer of $g_{\phi_2}$, which is defined as:
\begin{equation}
    (o_1,\cdots,o_K) = W^{(\phi_2)}_3(\mathrm{ReLU}(W^{(\phi_2)}_2(\mathrm{ReLU}(W^{(\phi_2)}_1 X))))
\end{equation}

Given the definition of $g_{\phi_1}$ and $g_{\phi_2}$, the objective loss function in PMT-IQA can be defined in Eq.~(\ref{eq::objective}).
\begin{equation}
\lambda_1 \mathcal{L}_r(x, y) + \lambda_2 \mathcal{L}_c(x, y)
\label{eq::objective}
\end{equation}
where $\mathcal{L}_r$ and $\mathcal{L}_c$ are the loss terms for the image quality score regression task and image quality level classification task, respectively. Parameters $\lambda_1, \lambda_2 > 0$ are dynamic hyper-parameters in the training procedure. In our implementation, we use $\ell_1$ loss (defined as Eq.~(\ref{eq:regression_loss})) for $\mathcal{L}_r$ and cross-entropy loss (defined as Eq.~(\ref{eq:cls_loss})) for $\mathcal{L}_c$, respectively.
\begin{equation}
    \mathcal{L}_r(x, y) = \frac{1}{n} \sum_{i=1}^{n}|y_i-g_{\phi_1}(h(x_i))|
    \label{eq:regression_loss}
\end{equation}
\begin{equation}
    \mathcal{L}_c(x, y) = -\frac{1}{n}\sum_{i=1}^{n}\sum_{c=1}^{K}(y_{i})_c\log((g_{\phi_2}(h(x_i)))_c)
    \label{eq:cls_loss}
\end{equation}
where $(\cdot)_c$ in Eq.~(\ref{eq:cls_loss}) denotes the \textit{c}-th component of $(\cdot)$. 

To simulate the from easy to hard learning law~\cite{pavlov1927conditioned}, we make the model focusing on learning the classification task in the early stage of training, and gradually concentrates on scalar quality score assessment with the progress of training by dynamically adjusting the weights of the classification and regression tasks as:

\begin{equation}
    \begin{aligned}
        \lambda_1(t)&=\frac{t}{T+1}\xi \\ 
        \lambda_2(t)&=1-\lambda_1(t)
    \end{aligned}
\label{eq::hyperparameter}
\end{equation}

where $t$ represents the $t$-th epoch, $T$ denotes the maximum epochs. $\xi$ is a trade-off to balance the two losses' scale difference.
We adopt the Adam optimizer~\cite{kingma2014adam} to optimize the PMT-IQA parameters $\phi$ and $\psi$ jointly.

\section{Experiment}
\label{sec:pagestyle}

\subsection{Experimental Setup}
\subsubsection{Datasets}
We use four publicly available IQA datasets, including LIVE Challenge (LIVE-C)~\cite{ghadiyaram2015massive}, BID~\cite{ciancio2010no}, LIVE~\cite{sheikh2006statistical}, and CSIQ~\cite{larson2010most}, to evaluate each IQA method. In these four datasets, BID and LIVE-C are authentic distortion datasets, where BID contains 586 figures with realistic blurry distortions, and LIVE-C includes 1162 real-world images collected by various cameras. In addition to authentic distortion datasets, we also evaluate PMT-IQA on two synthetic image datasets LIVE and CSIQ, which contain 779 and 866 images with 5 and 6 individual distortions, respectively.

\subsubsection{Evaluation Metrics}
We select two commonly-used evaluation metrics, Spearman's rank-order correlation coefficient~(SRCC)~\cite{lehman2005jmp} and Pearson's linear correlation coefficient~(PLCC)~\cite{lehman2005jmp}, 
to evaluate the performances of IQA algorithms. The definition of SRCC and PLCC are presented in Eq.~\ref{eq:srcc} and Eq.~\ref{eq:plcc}, respectively. Both SRCC and PLCC range from -1 to 1, and a larger value indicates a better performance. 
\begin{equation}\label{eq:srcc}
    \mathrm{SRCC}=1-\frac{6\sum_i d_i^2}{n\left(n^2-1 \right)}
\end{equation}
\begin{equation}\label{eq:plcc}
    \mathrm{PLCC}=\frac{\sum_i \left(q_i-q_m \right)\left(\hat{q}_i-\hat{q}_m \right)}{\sqrt{\sum_i\left(q_i-q_m \right)^2\sum_i\left(\hat{q}_i-\hat{q}_m \right)^2}}
\end{equation}
where $d_i$ denotes the rank difference between MOS and the predicted score of the \textit{i}-th image, $n$ is the number of images. $q_i$ and $\hat{q}_i$ are MOS (DMOS) and the predicted score of the \textit{i}-th image respectively, and $q_m$ and $\hat{q}_m$ are average MOS (DMOS) value and average predicted score for all images.

\subsubsection{Implementation Details}
\begin{figure}[ht]
    \centering
    \includegraphics[width=0.65\linewidth]{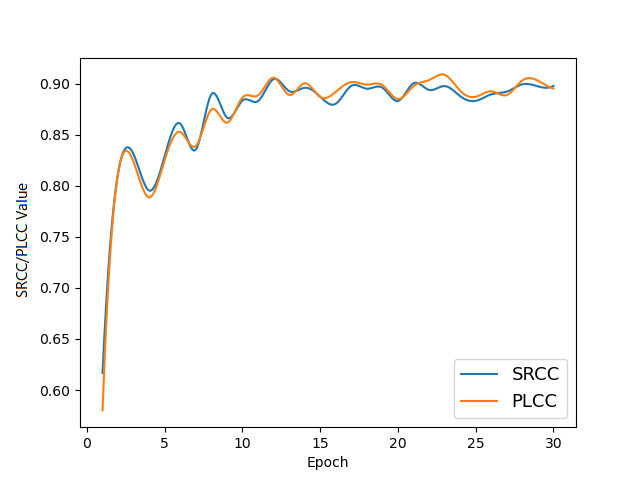}
    \caption{SRCC and PLCC values of PMT-IQA on BID dataset in the training procedure.}
    \label{fig::train_curve}
\end{figure}

We follow the experimental protocol and settings in HyperIQA~\cite{su2020blindly} for a fair comparison. Each dataset is divided into train and test set according to 4:1. The quality scores are scaled into [0,1] to improve stability, as shown in Fig.~\ref{fig::train_curve}.
During training, we augment each training image by randomly cropping and horizontally flipping ten times for LIVE-C and five times for the other three datasets according to HyoerIQA~\cite{su2020blindly}. A recently proposed hyperparameter searching framework optuna~\cite{akiba2019optuna} is employed to optimize hyperparameters and the values of hyperparameters of PMT-IQA on four datasets are reported in Table~\ref{tab::hyper}. In addition, dropout~\cite{srivastava2014dropout}
and weight-decay~\cite{krogh1991simple}
strategies are used to avoid over-fitting.

\begin{table}[ht]
\centering
\caption{The hyperparameters obtained by Optuna on the four test datasets.}
\begin{adjustbox}{width=0.5\textwidth}
\begin{tabular}{ccccc}
\toprule
Dataset  & LR          & Batch   & $\xi$  & Optimizer\\ \hline
BID     & 1.09e-4      &   12    &  0.9419       & Adam       \\
LIVE-C  & 4.72e-4      &   12    &  0.9841       & Adam      \\
LIVE    & 3.23e-4      &   12    &  0.9941       & Adam     \\
CSIQ    & 4.72e-4      &   12    &  0.8931       & Adam      \\  \bottomrule
\end{tabular}
\end{adjustbox}
\label{tab::hyper}
\end{table}

\subsection{Performance Evaluation}
We select fourteen representative BIQA methods as strong baselines, including BRISQUE~\cite{mittal2012no}, ILNIQUE~\cite{zhang2015feature}, AlexNet~\cite{krizhevsky2012imagenet}, ResNet50~\cite{he2016deep}, HOSA~\cite{xu2016blind}, BIECON~\cite{kim2016fully}, SFA~\cite{li2018has}, PQR~\cite{zeng2018blind}, DB-CNN~\cite{zhang2018blind}, HyperIQA~\cite{su2020blindly}, UNIQUE~\cite{zhang2021uncertainty}, CONTRIQUE~\cite{madhusudana2022image}, GraphIQA~\cite{sun2022graphiqa}, and TRes~\cite{golestaneh2022no}, to evaluate the performance of our proposed PMT-IQA model.

The SRCC and PLCC values of each method on the four test datasets are listed in Table~\ref{tab:overall-results}. From Table~\ref{tab:overall-results}, we can find that the PMT-IQA approach outperforms all the comparison methods on BID and LIVE-C for both SRCC and PLCC evaluation. Our PMT-IQA model achieves the second-best results on LIVE dataset, only weaker than GraphIQA. For the other synthetic distortion dataset CSIQ, PMT-IQA also obtains the best SRCC value (0.949) and competitive PLCC value (0.951). 
 In order to compare the performance of each method more intuitively, we also provide the ranks of all methods (i.e. the numbers in parentheses in Table~\ref{tab:overall-results}) and the average ranks of SRCC and PLCC of each method (i.e. the last two columns of Table~\ref{tab:overall-results}). The average SRCC and PLCC ranks of the proposed PMT-IQA on the four test datasets are 1.25 and 2.00, respectively, which are much better than the state-of-the-art methods GraphIQA~\cite{sun2022graphiqa} (average SRCC rank is 3.50, and 3.75 for average PLCC rank) and TReS~\cite{golestaneh2022no} (average SRCC rank is 4.00, and 3.75 for average PLCC rank).  All these results indicate tha our proposed PMT-IQA model obtains the best overall performance compared to the comparison methods.

\begin{table*}[ht]
    \centering
    \caption{The SRCC and PLCC values of various methods on BID, LIVE-C, LIVE and CSIQ datasets and the average rank of SRCC and PLCC for each method. Best performance in boldface and numbers in parentheses indicate corresponding ranks. We report the median SRCC and PLCC in  ten runs.}
    \small
    \begin{adjustbox}{width=1.0\textwidth}
    \begin{tabular}{c|cccc|cccc|cc}
    \toprule
       & \multicolumn{2}{c}{BID}   & \multicolumn{2}{c}{LIVE-C}  & \multicolumn{2}{c}{LIVE} & \multicolumn{2}{c}{CSIQ} & \multicolumn{2}{c}{\textit{Average Rank of }}\\ \hline
    Methods & SRCC & PLCC &  SRCC & PLCC & SRCC & PLCC & SRCC & PLCC & SRCC & PLCC\\
    \toprule
    BRISQUE~\cite{mittal2012no}  & 0.562(11) & 0.593(11) & 0.608(9) & 0.629(13) & 0.939(12) & 0.935(11) & 0.746(14) & 0.829(10) & 11.50(13) & 11.25(12) \\
    AlexNet~\cite{krizhevsky2012imagenet}  & -    &   -  & 0.766(10) & 0.807(11) & 0.932(13) & 0.841(15) & 0.766(13) & 0.811(13) & 12.00(14) & 13.00(14) \\
    ResNet50~\cite{he2016deep} & 0.583(10) &  0.599(10) & 0.824(9) & 0.868(7) & 0.947(10) & 0.913(12) & 0.823(9)  & 0.876(9) & 9.50(9) & 9.50(9) \\
    ILNIQE~\cite{zhang2015feature}   & 0.516(13) & 0.554(13) & 0.432(11) & 0.508(15) & 0.903(14) & 0.865(14) & 0.806(11) & 0.808(14) & 12.25(15) & 14.00(15) \\
    HOSA~\cite{xu2016blind}     & 0.721(9) & 0.736(9) & 0.640(8) & 0.678(12) & 0.946(11) & 0.947(10) & 0.741(15) & 0.823(11) & 10.75(11) & 10.50(10) \\
    BIECON~\cite{kim2016fully}   & 0.539(12) & 0.576(12) & 0.595(10) & 0.613(14) & 0.961(8) & 0.962(7) & 0.815(10) & 0.803(15) & 10.00(10) & 12.00(13)\\
    SFA~\cite{li2018has}      & 0.826(7) & 0.840(7) & 0.812(10) & 0.833(10) & 0.883(15) & 0.895(13) & 0.796(12) & 0.818(12) & 11.00(12) & 10.50(10) \\
    PQR~\cite{zeng2018blind}      & 0.775(8) & 0.794(8) & 0.857(3) & 0.872(5) & 0.965(6) & 0.951(9) & 0.873(8) & 0.901(8) & 6.25(7) & 7.50(8) \\
    DB-CNN~\cite{zhang2018blind}   & 0.845(6) & 0.859(6) & 0.851(5) & 0.869(6) & 0.968(5) & 0.971(2) & 0.946(3) & \textbf{0.959}(1) & 4.75(6) & 3.75(2) \\
    HyperIQA~\cite{su2020blindly}  & 0.869(3) & 0.878(3) & 0.859(2) & 0.882(3) & 0.962(7) & 0.966(6) & 0.923(5) & 0.942(5) & 4.25(4) & 4.25(5) \\
    UNIQUE~\cite{zhang2021uncertainty} & 0.858(5) & 0.873(4) & 0.854(4) & 0.890(2) & 0.969(2) & 0.968(4) & 0.902(7) & 0.927(7) & 4.50(5) & 4.25(5) \\
    CONTRIQUE~\cite{madhusudana2022image} & - & - & 0.845(7) & 0.857(9) & 0.960(9) & 0.961(8) & 0.942(5) & 0.955(3) & 7.00(8) & 6.67(7) \\
    GraphIQA~\cite{sun2022graphiqa} & 0.860(4) & 0.870(5) & 0.845(7) & 0.862(8) & \textbf{0.979}(1) & \textbf{0.980}(1) & 0.947(2) & \textbf{0.959}(1) & 3.50(2) & 3.75(2) \\ 
    TReS~\cite{golestaneh2022no} & 0.872(2) & 0.879(2) & 0.846(6) & 0.877(4) & 0.969(2) & 0.968(4) & 0.922(6) & 0.942(5) & 4.00(3) & 3.75(2)\\
    \hline
    PMT-IQA (Proposed) & \textbf{0.874}(1) & \textbf{0.894}(1) & \textbf{0.866}(1) &  \textbf{0.893}(1)   & 0.969(2) & 0.971(2)  & \textbf{0.949}(1) & 0.951(4) & \textbf{1.25}(1) & \textbf{2.00}(1) \\
    \bottomrule
    \end{tabular}
    \end{adjustbox}
    \label{tab:overall-results}
\end{table*}

To verify the generalization ability of the proposed PMT-IQA model, we further conducted several cross-database tests. In this test, a specific dataset is used as a training set and a different dataset plays the role of test set. We use four competitive methods, including PQR, DB-CNN, HyperIQA and TReS, as baselines for performance comparison. The cross-database tests include four settings: (1) train on LIVE-C and test on BID, (2) train on BID and test on LIVE-C, (3) train on LIVE and test on CSIQ, and (4) train on CSIQ and test on LIVE. The first two settings are for authentic distortion, while the last two are for synthetic distortion. Table~\ref{tab:crossdatabase} presents the SRCC values of the comparison methods and our proposed PMT-IQA model. We can see from the results that the proposed PMT-IQA model significantly outperforms the compared methods on all four cross-database tests.

\begin{table}[htp]
\centering
\caption{SRCC comparison on cross-database tests. Best results in boldface.}
\begin{adjustbox}{width=1.0\textwidth}
\begin{tabular}{cc|cccc|c}
\toprule
Train on & Test on &  PQR~\cite{zeng2018blind}&  DB-CNN~\cite{zhang2018blind} &  Hyper-IQA~\cite{su2020blindly} & TReS~\cite{golestaneh2022no} & PMT-IQA (Proposed) \\ \hline
LIVE-C  &  BID      &   0.714   &  0.762  & 0.756  & 0.870 &  \textbf{0.897} \\
BID     &  LIVE-C   &   0.680   &  0.725  & 0.770  & 0.765 &  \textbf{0.782} \\
LIVE    &  CSIQ      &  0.719   &  0.758  & 0.744  & 0.738 &  \textbf{0.766} \\
CSIQ    &  LIVE      &  0.922   &  0.877  & 0.926  & 0.932 &  \textbf{0.934}\\ 
\bottomrule
\end{tabular}
\end{adjustbox}
\label{tab:crossdatabase}
\end{table}

\subsection{Ablation Study}
We conduct several subtle ablation studies on the four test datasets to further verify the effectiveness of MS and PMT modules. The variants are as follows:
\begin{enumerate}
    \renewcommand{\labelenumi}{(\theenumi)}
    \item \textbf{ResNet}: Pre-trained ResNet50 architecture on ImageNet, adding fully connected layers for prediction~(i.e., without  MS and PMT).
    \item  \textbf{Type1}: The entire architecture in Fig.~\ref{fig::architecture} with only MS~(i.e., without PMT).
    \item  \textbf{Type2}: The entire architecture in Fig.~\ref{fig::architecture} with MS and PMT using fixed $\lambda_1$ and $\lambda_2$, and we use $\lambda_1=\lambda_2=0.5$ in our implementation based on test experiments.
    \item \textbf{PMT-IQA}: The entire architecture PMT-IQA in Fig.~\ref{fig::architecture} with MS and PMT 
using dynamic task weights as Eq.~\ref{eq::hyperparameter}.
\end{enumerate}

\begin{figure*}[t]
    \centering
    \includegraphics[width=0.71\textwidth]{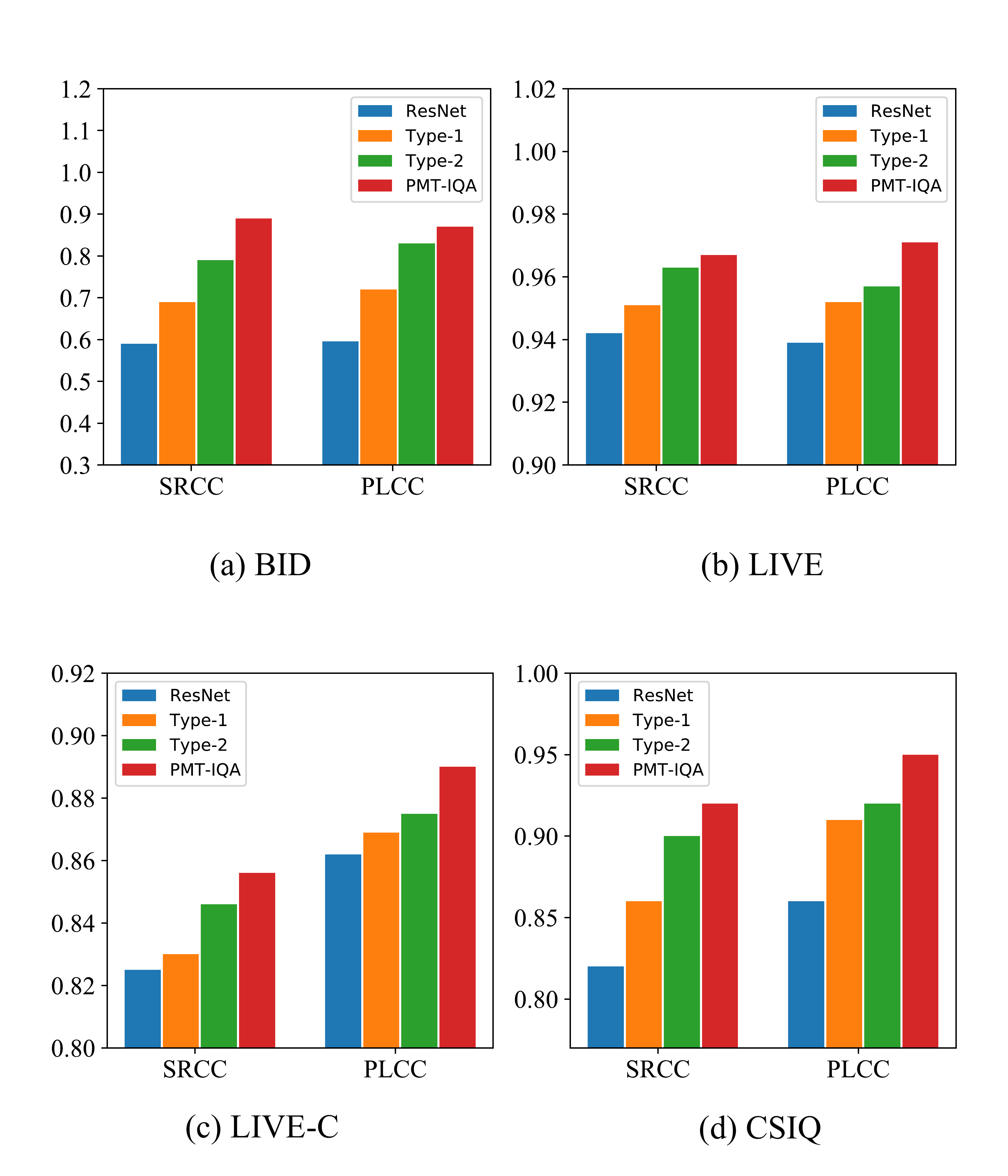}
    \caption{The ablation study on the BID, LIVE, LIVE-C and CSIQ datasets.}
    \label{fig::ablation}
\end{figure*}

We tune the hidden dimension to ensure variants have similar numbers of parameters to the completed PMT-IQA by removing the performance gain induced by model complexity for fairness.
As shown in Fig.~\ref{fig::ablation}, we can see that PMT-IQA achieves the best performance compared with the other three variants on all four test datasets. 
In addition, both MS and PMT modules can bring performance improvement to the model. Compared with MS, PMT improves the overall performance more significantly for PLCC evaluation on LIVE, LIVE-C and CSIQ. Moreover, the comparison between the results of Type2 and PMT-IQA shows that the strategy of dynamically adjusting the task weights to make the network learn from an easy task to a complex task is effective.
The novel progressive shift of tasks in PMT-IQA is essential in the training stage. 
Therefore, the ablation study results again verify the effectiveness of the proposed PMT-IQA.

\section{Conclusion}
\label{sec:typestyle}

In this paper, we propose a simple yet effective progressive multi-task learning model for blind image quality assessment. Our model contains a multi-scale feature extraction module and a progressive multi-task learning module to help the model learn complex distortion patterns by simulating the from easy to hard human learning law. Extensive experimental results show that 
despite the relatively simple architecture of the proposed PMT-IQA method, it can still achieve superior or competitive performance compared to various baselines on all the datasets.

\subsubsection*{Acknowledgments.}
This work is supported by the National Natural Science Foundation of China under Grant 61866031.



%
%
%
%

\newpage
\appendix


\end{document}